\documentclass[12pt,preprint]{aastex}

\shorttitle{Far Ultraviolet Continuum Emission of Solar-Mass Stars}
\shortauthors{Linsky et al.}
\slugcomment{Draft \today}

\begin{document}

\newcommand{\php}[0]{\phantom{--}}
\newcommand{\kms}[0]{km~s$^{-1}$}

\title{FAR ULTRAVIOLET CONTINUUM EMISSION: APPLYING THIS DIAGNOSTIC TO THE 
CHROMOSPHERES OF SOLAR-MASS STARS}

\author{Jeffrey L. Linsky\altaffilmark{1}}
\affil{JILA, University of Colorado and NIST, 440UCB Boulder, CO 80309-0440, 
USA}
\email{jlinsky@jilau1.colorado.edu}

\author{Rachel Bushinsky}
\affil{APS, University of Colorado, 391UCB Boulder, CO 80309-0391, USA}

\author{Tom Ayres\altaffilmark{1}}
\affil{CASA, University of Colorado, 593UCB Boulder, CO 80309-0593, USA}

\author{Juan Fontenla}
\affil{LASP, University of Colorado, 590UCB Boulder, CO 80309-0590, USA}

\and

\author{Kevin France}
\affil{CASA, University of Colorado, 593UCB Boulder, CO 80309-0593, USA}

\altaffiltext{1}{Guest Observer, NASA/ESA {\it Hubble Space Telescope} and 
User of the Data Archive at the Space Telescope 
Science Institute. STScI is operated by the Association of Universities for 
Research in Astronomy, Inc., under NASA contract NAS 5-26555. These 
observations were made as parts of programs \#11532, \#11534, and \#11687.}

\begin{abstract}

\noindent The far ultraviolet (FUV) continuum flux is recognized as a very 
sensitive diagnostic of the temperature structure of the Sun's lower 
chromosphere. Until now analysis of the available stellar FUV data has shown  
that solar-type stars must also have chromospheres, but quantitative 
analyses of stellar FUV continua require far higher quality spectra and
comparison with new non-LTE chromosphere models. We present accurate far 
ultraviolet (FUV, 1150--1500~\AA) continuum flux measurements for solar-mass 
stars, made feasible by the
high throughput and very low detector background of the Cosmic Origins 
Spectrograph (COS) on the {\em Hubbble Space Telescope}. 
We show that the continuum flux can be measured
above the detector background even for the faintest star in our sample. 
We find a 
clear trend of increasing continuum brightness temperature at all FUV 
wavelengths with decreasing rotational period,
which provides an important  measure of magnetic heating rates in 
stellar chromospheres. Comparison with semiempirical solar flux models
shows that the most rapidly rotating solar-mass stars have FUV continuum
brightness temperatures similar to the brightest faculae seen on the Sun. The 
thermal structure of the brightest solar faculae therefore provides a 
first-order estimate of the thermal structure and heating rate 
for the most rapidly rotating solar-mass stars in our sample. 

\end{abstract}


\section{INTRODUCTION}

Above the solar photosphere, the conversion of magnetic and wave energies into 
heat produces a thermal inversion called the chromosphere.
Stars with spectral types later than about A7~V have convective zones, 
magnetic dynamos, and chromospheres, although the heating processes
may change with stellar effective temperature and magnetic field structure. 
The thermal structures and energy budgets of different regions of the solar
chromosphere have been studied by analysis of emission
lines such as the Ca~II H and K lines and the  
far ultraviolet (FUV) continuum flux. Semiempirical solar chromosphere 
models, such as 
those computed by \citet{Vernazza1976}, \citet{Vernazza1981}, 
\citet{Avrett2008}, and \citet{Fontenla2009} are 
based in large part on fitting the observed FUV continuum.

Until recently, it has not been possible to measure accurate 
FUV continuum fluxes of solar-type stars because of their faint signal 
and large instrumental background noise produced by detector dark current and 
scattered light. For example, \citet{Franchini1998} showed that scattered 
light below 1500~\AA\ is a large portion of the signal obtained by the 
International Ultraviolet Explorer (IUE) satellite, although they could 
approximately correct the data for scattered light. Also detector background
is clearly present in the far-UV as the contrast between emission lines and
the flux between the emission lines is often very low. 
As a result, chromosphere models of solar-type stars
\citep[e.g.,][]{Vieytes2005} are typically based on fitting
only the Ca~II H and K and hydrogen Balmer emission lines. 
However, these emission lines are formed at the base of the 
chromosphere, whereas in the Sun the FUV continuum flux at wavelengths 
shortward of 1500~\AA\ 
is formed in, and is diagnostic of, the chromospheric temperature rise
from the temperature minumum. 

\citet{Franchini1998} showed that G-star photospheric models with a monotonic 
decrease in temperature with height severely underpredict the observed
UV flux below 2000~\AA. The addition of an extended temperature minimum 
structure with an empirical value of $T_{\rm min}$ provides better fits to the 
spectral flux of G-type stars between 1600 and 2000~\AA, but such models 
severely underpredict the flux at shorter wavelengths
\citep{Franchini1998,Morossi2003}. 
To fit the observed FUV flux below 1600~\AA\ requires temperatures rising with 
height above the temperature minimum -- a chromosphere. 
Using a semi-empirical solar chromospheric model, 
\citet{Morossi2003} could approximently fit the FUV spectrum
of HD~131156 ($\xi$ Boo A).

\citet{Franchini1998} and \citet{Morossi2003} thus demonstrated
that the FUV flux below 1600~\AA\ requires that solar type 
stars have solar-like chromospheres. Unfortunately, the available but rather 
noisy IUE data could not
provide accurate measurements of the far-UV continuum uncontaminated
by detector background and weak emission lines and only a few higher-quality 
by low-resolution HST/GHRS G140L spectra were available for analysis. 
Their comparison of the $\xi$~Boo~A spectrum with scaled solar-models 
including a chromosphere temperature structure was also uncertain
as the spectral fitting software assumed LTE and the bound-free opacity of
neutral metals that dominates the far-UV continuum depends on non-LTE
processes. As a result, they were not able to find any trends in the far-UV 
emission with stellar rotation rate or other indicators of stellar magnetic 
activity.

What has been lacking in the previous studies is the availability of 
accurate FUV continuum fluxes of solar-type stars with a range of magnetic 
activity that can be fitted with semi-empirical non-LTE models of stellar
chromospheres. We show in this paper that the combination of high
throughput and very low detector background of the Cosmic Origins Spectrograph
(COS) on HST now permits FUV continuum measurements for many
stars. We analyze here COS observations of six stars with masses close to 
solar, but with a wide range of rotational periods.
 
Solar-mass stars reach the zero-age main sequence after their premain 
sequence evolution as rapid rotators with strong magnetic fields. 
As they age on the main sequence, the torque of their magnetized winds slows
their rotation \citep[e.g.,][]{Matt2008}, producing many profound
changes in their magnetic structure 
and emission properties. \citet{Wilson1963} and \citet{Kraft1967},
among others, showed that the strength of the chromospheric Ca~II lines 
decreases with age and rotational velocity for solar-type stars. 
Other changes also occur in the X-ray \citep[e.g.,][]{Pizzolato2003},
UV \citep[e.g.,][]{Simon1985}, and radio emission, starspot properties, 
and magnetic field
structure. These changes with rotation suggest that the FUV continuum
should also depend on rotation rate, but this suggestion has until now not 
been properly tested.

\section{{\em HST} OBSERVATIONS OF SOLAR-MASS STARS}

Our sample of solar-mass stars extending over a range of rotation periods 
and ages (see Table~1) was observed with the G130M grating on COS
as portions of three {\em Hubble Space Telescope} programs. For a
description of the COS instrument and its on-orbit performance,  
see \citet{Osterman2011}. The resolving
power of the COS G130M grating is $R\approx$ 17,000--18,000.
We processed the data with the COS calibration pipeline, 
CALCOS\footnote{We refer the reader to the cycle 18 COS Instrument Handbook 
for more details: 
http://www.stsci.edu/hst/cos/documents/handbooks/current/cos\_cover.html}
v2.11, combined with a custom IDL coaddition procedure. 
Table~1 provides details of these observations, and Figure~1 shows portions 
of these spectra.

The fastest rotating star in our sample is the Pleiades Cluster star 
\object{HII314}
(G1~V) with a rotation period $P_{\rm rot}=1.27$ days \citep{Rice2001}. 
We observed this star for 4611 seconds in program 11532 covering the 
spectral range $1134\leq \lambda \leq 1459$
with four central wavelength settings ($\lambda$1291, $\lambda$1300,
$\lambda$1309, and $\lambda$1318) to minimize fixed-pattern noise in the
detector.

\object{HD~209458} is the G0~V host star of the extensively studied 
exoplanet \object{HD~209458b}. The rotational period $P_{\rm rot}= 11.4$ days  
was measured by \citet{Silva-Valio2008} from occultations of starspots 
by its transiting planet. We observed \object{HD~209458} for a total of 
25641 seconds in program 11534 during transit, secondary eclipse, and both 
quadratures, as described by \citet{France2010a} and
\citet{Linsky2010}. We analyze here the combined 
data outside of transit from the same four central wavelength settings used 
for \object{HII314} including the wavelength range 1140--1450~\AA. 
We also include the G160M observations described by \citet{France2010a}
to extend the wavelength range to 1500~\AA.

\object{EK Dra} (G1~V), \objectname[Pi1 UMa]{$\pi^1$ UMa} (G1.5~V), 
\objectname[Chi1 Ori]{$\chi^1$ Ori} (G1~V), and \object{HD~25825} (G0~V)
were observed as part of SNAP program 11687 with partial orbit observations 
at the $\lambda$1291 central wavelength setting and two FP-POS positions
(3\&4) to mitigate fixed patterns. These observations cover the
wavelength region 1290--1430~\AA. The results of this SNAP program, with the 
objective to study the Fe~XXI 1354~\AA\ coronal emission line, will be 
presented elsewhere, although \citet{Ayres-France2010} have discussed the 
\object{EK Dra} emission line data.
Exposure times for the four stars in the SNAP program selected for this study 
were between 1160 and 1300 seconds. These data were reduced in the same way 
as described for \object{HII314}. \object{EK Dra}
is a member of the Pleiades Moving Group, \objectname[Pi1 UMa]{$\pi^1$~UMa} 
and \objectname[Chi1 UMa]{$\chi^1$~Ori} are
members of the Ursa Major Moving Group, and \object{HD~25825} is a member of 
the Hyades 
Cluster. Rotational velocities for these stars are listed in Table~1.
The ages listed in Table~1 are from \citet{Ribas2005} and \citet{Barnes2007}.

We include for comparison the very high-resolution ($R=114,000$)
STIS E140H spectrum of \objectname[Alpha Cen A]{$\alpha$~Cen~A} (G2~V) 
\citep{Pagano2004}, a slowly rotating near twin of the \object{Sun}. 
The data were taken from the StarCAT STIS spectral catalog \citep{Ayres2010}.
Although the detector noise background of the 
FUV channel of STIS is much larger than for COS, the strong signal of this 
nearby star makes FUV continuum measurements feasible.

\subsection{Comparison Solar Irradiance Measurements}

We also compare the FUV continuum fluxes of the solar-mass stars with 
corresponding solar irradiance measurements, which are flux values of 
the Sun viewed as an unresolved source. We use the solar irradiance 
reference spectrum obtained with the Solar Radiation
and Climate Experiment (SORCE) on the Solar-Stellar Irradiance Comparison 
Experiment II (SOLSTICE II) \citep{Woods2009, McClintock2005, Snow2005}.
These data cover the 1150--3200~\AA\ spectral range with 1~\AA\ resolution. We
selected these data because the absolute flux calibration is accurate to
about 5\% in the FUV and is cross-checked against B and A-type stars. Scattered
and stray light were removed from this data set. We use here the 2008 April 
10--16 irradiances that represent the Sun very close to minimum with a Zurich 
sunspot number of 2 and an average 10.7 cm radio flux of $68.9\times 10^{-22}$ 
Wm$^{-2}$Hz$^{-1}$.

\subsection{Measurement of the COS Instrumental Background}

There are three sources of background that must be evaluated before we
can reliably ascribe the measured fluxes to stellar FUV continuum emission. 
First, the Exposure Time 
Calculator\footnote{http://etc.stsci.edu/etc/input/cos/spectroscopic/} 
(ETC) for the COS instrument estimates that the accumulated FUV sky background 
is below $10^{-3}$ counts over the duration of each COS observation and 
can therefore be ignored.
Second, prior to launch, \citet{Osterman2002} measured the amount of light
scattered into a 1 \AA\ bandpass by the G130M grating to be less than
$<2\times 10^{-5}$ of all light within $\pm 10$~\AA.
This very low value rules out any significant contribution to the background,
except perhaps within 10~\AA\ of the Lyman-$\alpha$ line center. 
The on-orbit scattered-light 
level has been estimated to be twice as large, but this will also not be 
significant. Finally, detector noise is the most important background.
The on-orbit-measured detector count rate is $\sim 1.8\times 10^{-6}$ counts
s$^{-1}$ pixel$^{-1}$, although the ETC assumes a more conservative value of
$\sim 2.25\times 10^{-6}$ counts s$^{-1}$ pixel$^{-1}$. The estimated detector
background counts for all of the measurements using the on-orbit count rate
are listed in Table 2.

Since the detector dark count rate dominates the
background signal, we also measure it
by a different method as an independent check.
Following the procedure outlined in \citet{France2010b}, we use the time-tag
capability of the COS microchannel plate (MCP) detector to measure
the relative contributions of the FUV continuum and background rates,
independent of the standard background subtraction applied by the CALCOS
pipeline to the one-dimensional science data. For each
target, we extract the location-time photon list [$x_{i}$,$y_{i}$,$t_{i}$] 
from each
exposure $i$ and coadd them into a master [$x$,$y$,$t$] list, taking into
account the appropriate pixel offsets for exposures made with different
central wavelength and FP-POS settings.  For the G130M A and B segments,
$\Delta\lambda$~=~[1339.8~--~1347.8~\AA] and [1244.8~--~1253.3~\AA],
respectively, we define a detector
[$\Delta$$x$,$\Delta$$y$] location corresponding to the wavelength region of
interest ($\Delta\lambda$)  and the active science area in the
cross-dispersion direction.  

We then integrate the total number of counts in the
[$\Delta$$x$,$\Delta$$y$] box in a timestep $\Delta$$t$ to compute
the count rate in a given spectral window as a function of time.  We use a
timestep of  $\Delta$$t$ = 200~s for this analysis.  The instrument
background level is computed in a similar manner, with the background
integrated over the same $\Delta\lambda$ as the continuum region, but offset
below the active science region 
($\Delta$$y_{\rm back}$ = $\Delta$$y_{\rm sci}$ - 50 pixels). 
We show an illustration of the continuum and background count
rates (binned by a factor of 2 for display purposes) for \object{HD 209458} 
in Figure 2.  For reference, we also show the count rate of the chromospheric
\ion{C}{2} 1334+1335~\AA\ emission lines measured with the same method.  
Because the \object{HD 209458} observing campaign extended over several 
weeks in 2009 September and
October, we plot the count rates as a function of total-integrated observing
time as opposed to absolute observing time.  We have not included 3,200 s of 
data when the background was high probably due to the spacecraft being near 
the South Atlantic Anomaly. One sees that the FUV
continuum level is significantly above the instrumental background at almost
all times over the 18 individual exposures ($T_{\rm exp}$~=~22,441 s) that went
into the master [$x$,$y$,$t$] photon list.  Summing over the total exposure
time, this method yields an integrated detection significance ($\sigma$), 
defined as the square root of the continuum minus background counts, of
$\sigma\sim 42$ for the continuum in the example shown in Figure 2. 
Correlation of the continuum and background signals has a Spearman rank
correlation index of 0.19 indicating no significant correlation.
Applying this technique to all of the stars in our sample, 
we detect far-UV continua at $\sigma> 12$ in all targets observed by COS. 
Systematic errors are discussed in the next section.

\section{RESULTS}

Since the COS spectra have very low detector background noise,
we can, for the first time, measure the FUV continuum flux between the 
emission lines. We have identified several 4--14~\AA\ wide windows with
no obvious emission lines in the COS spectra (see Figure~1) 
and only very weak or no emission lines in the very deep 
\objectname[Alpha Cen A]{$\alpha$~Cen~A} (G2~V) spectrum \citep{Pagano2004}. 
The flux axes in Figure~1 have been set to identify weak emission lines and 
the flux between the strong emission lines. 
We convert the average fluxes in these windows 
(units: ergs cm$^{-2}$ s$^{-1}$ \AA$^{-1}$) in Table~2 to 
stellar surface fluxes using the stellar distances and radii (and their 
uncertainties) listed in 
Table~1 and then to equivalent blackbody brightness
temperatures ($T_{\rm B}$). Figure~3 shows these brightness temperatures 
in the various windows for the six stars observed with COS, 
the STIS spectrum of $\alpha$~Cen~A, and the Sun (irradiance). 
For each spectral window, we also list in Table~2 the number of continuum 
and detector dark counts (using the on-orbit count rate) computed 
by the ETC. We assume that the measurement errors add in quadrature, that is,
the percent measurement errors listed in Table~2 are computed from the 
square root of the sum of the continuum and dark counts divided by the
continuum counts.

There are two other important sources of uncertainty - conversion of observed 
flux to surface flux and instrument calibration. These uncertainties can
be considered systematic as they apply to all data for each star. 
The ratio of surface to
observed flux is proportional to $(d/R)^2$. Table~1 lists the sources of
these quantities. Since most of the radii are not accurately measured, we
assign generous uncertainties. 

As described in the Cycle 19 COS Instrument
Handbook Section 6.1.8 and by \citet{Massa2010}, 
the absolute flux calibration of COS is accurate 
to about 5\% in the FUV. Also, time-dependent sensitivity corrections should 
be accurate to about 2\%. 
The absolute solar fluxes obtained with the SORCE instrument on SOLSTICE~II 
are accurate to about 5\% (see above). The absolute flux
uncertainty for the STIS E140H spectra depends on wavelength. 
\citet{Bohlin1998} finds that for the 1325--1510~\AA\ region, the uncertainty
is about 4\% based on \object{BD+28D4211} as the calibration star, 
but at shorter wavelengths the uncertainty is 10--20\% because 
\object{BD+75D325} is the calibration star. 

Since these four systematic errors (stellar distance, stellar radius, 
absolute flux calibration, and time-dependent sensitivity correction)
are independent, we sum them in quadrature. Finally we sum the systematic
and measurement errors in quadrature to obtain the uncertainty in
the surface flux and $T_{\rm B}$ values listed in Table 2. 

At wavelengths much shorter than the blackbody emission peak,  
small changes in $T_B$ correspond to large 
changes in flux and the brightness temperatures 
are, therefore, relatively insensitive to the FUV flux errors. 
For example, at 1415~\AA\ a $\pm 20$\% error for the solar flux  
corresponds to $T_{\rm B} = 4611.6^{+38.5}_{-46.2}$ K, and at 1165~\AA\ the 
brightness temperature range is $T_{\rm B} = 5260.2^{+41.1}_{-49.5}$~K. 
Therefore, the 
large increases in $T_{\rm B}$ at all FUV wavelengths for rapidly rotating 
solar-mass stars compared to the Sun far exceed the flux errors.

\section{DISCUSSION}

\citet{Vernazza1976} evaluated the then available solar FUV 
continuum data for disk center (radiance) and flux from the entire Sun 
(irradiance) using Skylab measurements that were carried out at only a few 
locations on the
 solar disk during low solar activity conditions and have 
uncertain flux-calibration errors. They determined brightness 
temperatures in the 1300--1700~\AA\ range that cluster about 4500~K 
for both the disk center radiance and whole disk flux data. 

\citet{Fontenla2009} computed a grid of one-dimensional non-LTE model 
atmospheres to match the observed spectra between the extreme ultraviolet 
and 100~$\mu$m for several types of solar regions defined by their Ca~II K 
line brightness and for sunspot umbrae and penumbrae. Five of the models were 
computed to match the continuum from the relatively dark 
quiet-Sun inter-network to bright faculae. \citet{Fontenla2011} improved the 
transition regions between the chromosphere and corona
of these models and included more species whose 
photoionization and photodissociation opacities replace the {\em ad hoc} 
extra FUV opacity used in the \citet{Fontenla2009} models. They also 
computed two new models, one for very dark (in the Ca~II K line) regions of 
the inter-network and another for the very brightest facular areas seen on 
the Sun.

The current set of models describes the temperature and density structure 
as functions 
of height from the deep photosphere up to the transition region and corona. 
As with previous solar models by \citet{Vernazza1981} and by 
\citet{Fontenla1993}, 
the temperature decreases with increasing height in the photosphere 
reaching a temperature minimum. Above the minimum, the temperature increase 
in the chromosphere produces the observed FUV continua and 
lines. The minimum temperature is higher for the brighter models, but unlike 
the previous models, the Fontenla et al. (2009 and 2011) models have a 
temperature minimum that occurs at higher altitudes (corresponding to lower 
densities) for the darker regions. In the models for the quiet-Sun regions, 
the temperature minimum value is very low, which explains the CO infrared 
lines as well as other spectral features that were not explained by the 
previous models.

In the \citet{Fontenla2011} models, the FUV continuum radiation in the 
1300--1500~\AA\ 
range is produced almost entirely by recombination of Si II, Mg II, and Fe II, 
and the photospheric radiation is blocked by the photoionization continuum 
opacity of Si I (with other contributors as well) just below the temperature 
minimum region. The Si I opacity is largely reduced in the temperature minimum 
region by the over-ionization (with respect to LTE) caused by 
UV irradiation from the upper chromosphere; therefore the ionization is 
much higher than in LTE and is mainly determined by the illumination from the 
upper chromosphere. \citet{Fontenla2009} noted that the FUV continuum 
is the most sensitive diagnostic of the slope of the temperature rise in 
the upper chromosphere.

For each of the models, spectral radiances at 10 points from the solar 
disk center to the limb are combined to provide the irradiance of a "star" 
with an atmosphere consisting entirely of this model. 
We include in Figure~3 brightness temperatures corresponding to the 
irradiances for each of the \citet{Fontenla2011} models. 
We plot $T_{\rm B}$ at four wavelengths (1180.0, 1321.5, 1398.6, and 
1500.0~\AA) for models 1001 (quiet Sun internetwork), 1002 (quiet Sun 
network lane), 1003 (enhanced network), 1004 (plage), 1005 (facula), and 
1008 (very bright facula). These models correspond to the full range 
of solar activity and non-radiative heating rates, except for flares. 
The 2008 April SORCE data, which represents the Sun when it was close 
to minimum, lies near model 1002 at all four
wavelengths. The FUV brightness temperatures for $\alpha$~Cen~A, which is a 
somewhat more evolved and luminous G2~V star than the Sun, lie just below the 
Sun and generally below solar model 1001. 

In Figure~3 the $T_{\rm B}$ range between the Sun and the most rapidly 
rotating stars
(HII314 and EK Dra) is nearly the same as the range between the quiet Sun 
inner network and very bright faculae. We note that the stars that rotate
about 4 times faster than the Sun (HD~25825, $\chi^1$~Ori, and $\pi^1$~UMa)
lie well below the fastest rotators (EK~Dra and HII314).
HD209458, which rotates at twice the speed of the Sun, lies about halfway
between the moderately fast rotators and the Sun and $\alpha$~Cen~A. 
Figure~4 compares $T_{\rm B}$ for the 1382-1392~\AA\ band with the stellar 
rotation period, $P_{\rm rot}$. There is a clear trend of increasing 
$T_B$ with decreasing $P_{\rm rot}$ for the solar-mass stars. The figure 
shows both linear and quadratic least-squares fits to the data. The
second order curve provides a better fit to the data.

The increase in $T_{\rm B}$ to shorter wavelengths for the rapidly-rotating 
solar-mass 
stars is similar to that of the slowly-rotating Sun and the models for regions 
of different activity on the Sun. This is expected as the models for regions 
on the Sun with increasing activity have temperature structures  
with height that are similar in shape near and above the temperature minimum.
Increasing temperature and thus ionization lead to higher FUV continuum 
emission,
unless the opacity sources change dramatically. The agreement of the FUV
continuum fluxes of the rapidly-rotating solar-mass stars with the models
for the brightest regions on the Sun indicate that the models for active 
regions on the Sun provide first order approximations to the temperature 
minimum and chromospheric thermal structure of the rapidly-rotating stars.

The one-dimensional \citet{Fontenla2011} models also predict fluxes of 
emission lines formed above the photosphere,
but the thermal structure of the Solar chromosphere and higher layers
is far more complex than even a mixture of one-component models due to
absorption by overlying filament structures,  
time-dependent heating, and dynamics. 
Active stars presumably also have such properties.
The fluxes and profiles of optically thick emission lines formed in the 
chromosphere will likely be most affected by three-dimensional structures.
Optically thick resonance line profiles of O~I and C~II, therefore, may not be 
well fit by one-dimensional Solar models. 
A better test of our comparison of active stars to active regions on the Sun 
should be the lower opacity Si~IV 
1393~\AA\ emission line formed in the transition region between the 
chromosphere and corona. Figure~5 compares the stellar line fluxes with
those of the \citet{Fontenla2011} models. The stellar Si~IV fluxes are plotted 
against the 1382--1392~\AA\ stellar continuum flux, both at a distance of 
1~AU. For comparison, Si~IV fluxes are plotted against the 1398.6~\AA\ 
continuum fluxes for the models, also assuming a distance of 1~AU. Except for
HD~209458, the model predictions are close to the observations.   
The observations and models both show the same trend of increasing line 
and continuum flux with increasing stellar 
rotation and Solar activity.

\section{CONCLUSIONS}

We show that the high throughput and low detector noise of the COS permits the
detection of FUV continuum emission from solar-mass stars for the first time. 
The observed FUV
brightness temperatures obtained from the solar irradiance and from 
the somewhat more evolved G2~V star $\alpha$~Cen A are very similar between 
1150 and 1500~\AA, 
indicating that the absolute flux scales of HST and SORCE are
compatable. We find a clear trend of increasing FUV continuum 
brightness temperatures with more rapid rotation for the six solar-mass stars 
observed by COS, $\alpha$~Cen~A, and the Sun. 
Comparison with semiempirical models for different regions 
on the Sun shows that the FUV continuum brightness temperatures derived 
from the surface fluxes 
of the most rapidly rotating stars in our sample are similar to or slightly 
exceed those of the brightest facular regions on the Sun.
Since the FUV continuum flux is the most sensitive diagnostic of the thermal 
structure of the lower Solar chromosphere, we believe that
the now measurable FUV continuum fluxes of solar-mass stars and likely a 
much wider range of stars with convective zones will be very useful sensitive 
diagnostics of the thermal 
structures of their chromospheres. The broad range of possible 
applications of this new chromospheric diagnostic should be explored in 
future papers. In particular, the FUV continuum diagnostic should provide 
new information on the amount of magnetic heating in the chromospheres of 
rapidly rotating and thus active stars. This is important both for our 
understanding of stellar physical processes and because FUV photons 
strongly affect the atmospheric chemistry and heating of exoplanets through 
the excitation and dissociation of such molecules as H$_2$O, CH$_4$,
and CO$_2$.

\acknowledgements

This work is supported by NASA through grants NNX08AC146, NAS5-98043, 
and HST-GO-11687.01-A to the University of Colorado at Boulder. 
We thank Tom Woods
for providing the SORCE data and Steven Osterman for information on the 
COS calibration.

{\it Facilities:} \facility{HST (COS)}, \facility{HST (STIS)}. 
\facility{SIMBAD}

\begin{figure}
\includegraphics[angle=90, scale=0.7]
{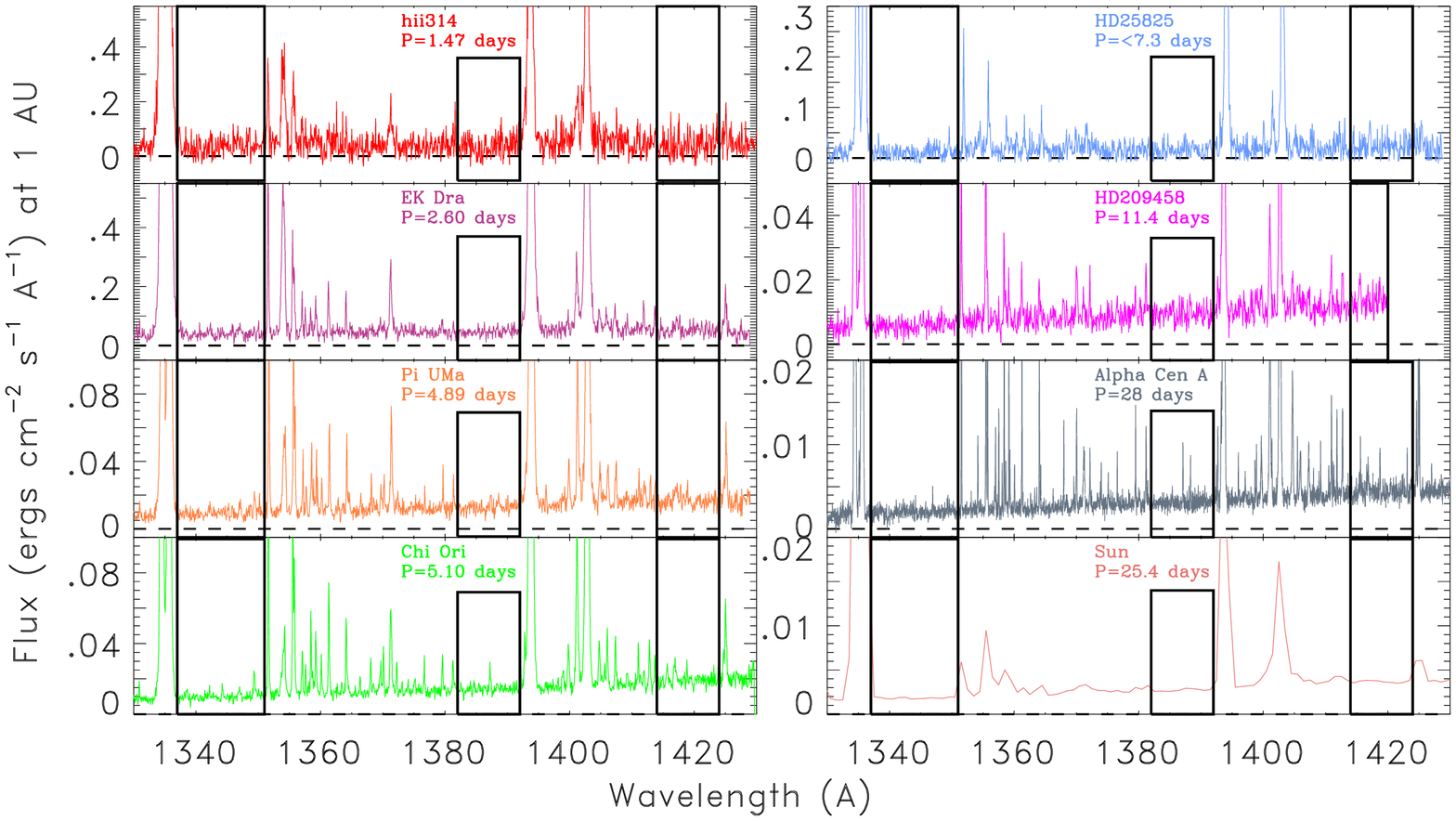}
\caption{Comparison of the spectra of the six solar-mass stars observed with 
COS, the STIS spectrum of $\alpha$~Cen~A, 
and the SORCE solar flux spectrum. 
The flux axes are set to identify weak emission lines and the stellar continua.
The centers of bright emission lines of C~II (1334 and 1335~\AA) and Si~IV
(1394 and 1403~\AA) are off scale.
The windows used for continuum measurements are indicated.}
\end{figure} 

\begin{figure}
\includegraphics[angle=90, scale=0.7]
{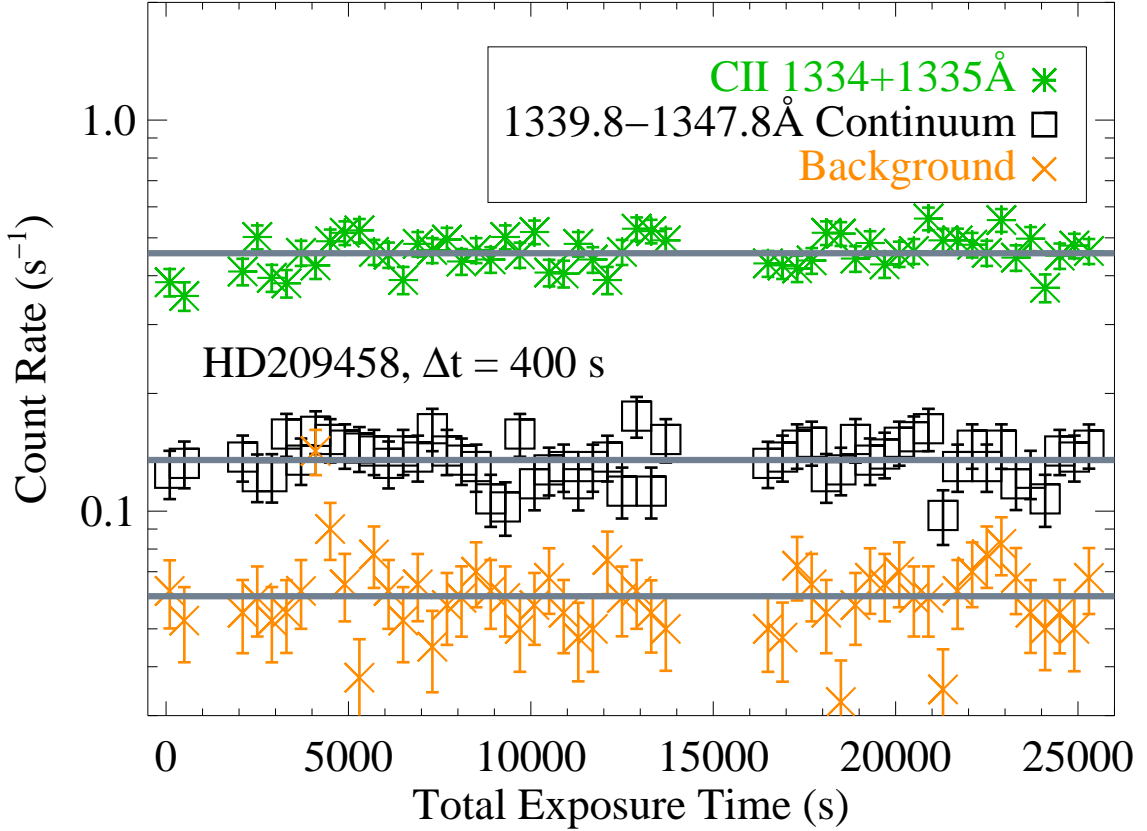}
\caption{The count rate of the G130M segment A continuum
(1339.8~$\lesssim$~$\lambda$~$\lesssim$~1347.8) in HD 209458 
(shown as black
squares) is compared with the detector background level measured immediately
below the active science area (shown as orange x's).  The coadded [x,y,t]
photon list was sampled at 200~s intervals, and the data have been rebinned by
a factor of two for display purposes. As the HD 209458 observing 
program spanned $\sim$~6 weeks, we show the count rates as a function 
of the total observing time. We have removed data in two time intervals of 
high background (800--1800 and 14000-16200 s) probably during spacecraft
passage near the South Atlantic Anomaly.    
The gray solid lines represent the average count rates.  For reference, the
count rate of the bright chromospheric emission lines from \ion{C}{2} 1334,
1335~\AA\ are shown as the green stars.}

\end{figure}

\begin{figure}
\includegraphics
{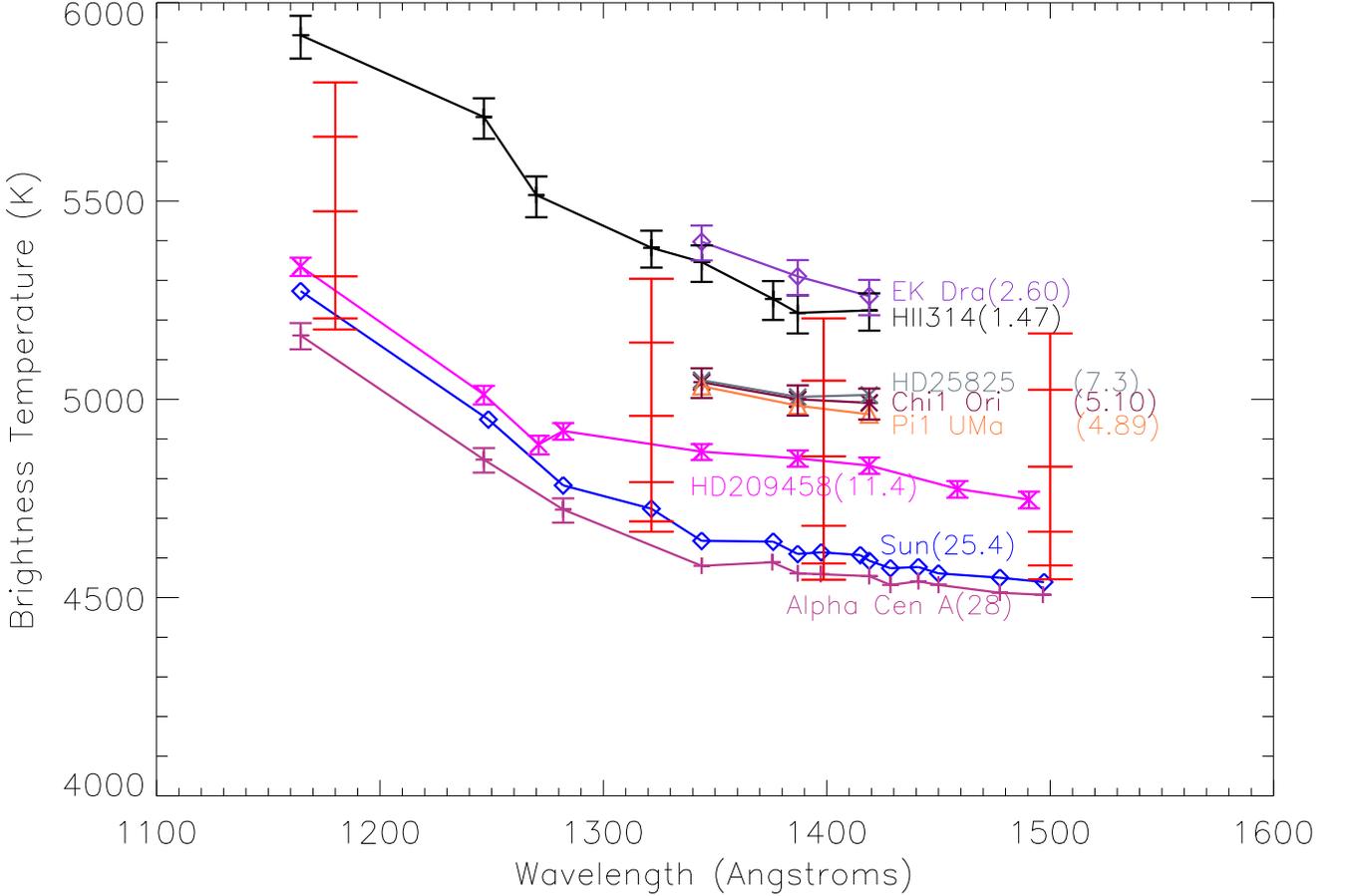}
\caption{Brightness temperatures vs. wavelength for the six solar-mass 
stars observed with COS, $\alpha$~Cen A 
observed with STIS, and the solar flux measured by
the SORCE instrument at a time of very low activity. 
The rotational periods of the stars and the 
siderial period of the Sun are in 
parentheses. Error bars computed from the measurement and systematic errors
listed in Table~2 are shown when the errors are greater than the 
symbol sizes. The error bars for $\chi^1$~Ori are also representative for
HD25825 and $\pi^1$~UMa.
The horizontal bars centered at wavelengths 1180, 
1321.5, 1398.6, and 1500.0~\AA\ indicate the predicted
brightness temperatures for 6 solar irradiance models computed by 
\citet{Fontenla2009} and \citet{Fontenla2011} for 
the Quiet Sun inner network (lowest bar) up to a 
very bright facula (Models 1001 to 1005 and 1008).}
\end{figure}

\begin{figure}
\includegraphics{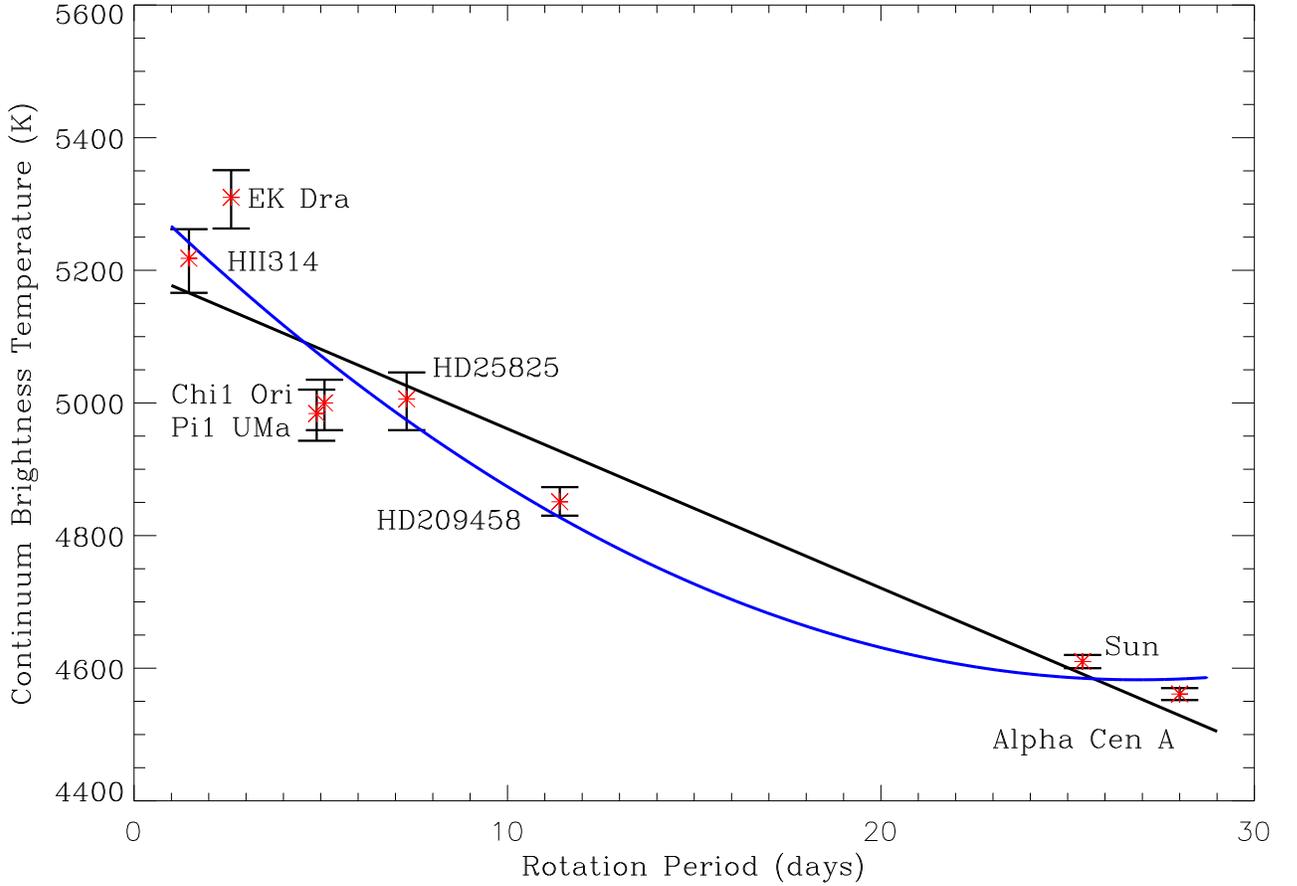}
\caption{Comparison of the 1382-1392~\AA\ continuum brightness temperatures
with rotation periods for the identified stars. The black line is a 
least-squares linear fit to the data, 
$T_{\rm B}=5201.14 - 24.0148P_{\rm rot}$. 
The blue line is a least-squares quadratic fit to the data,
$T_{\rm B}=5320.17 - 54.8303P_{\rm rot} + 1.01890P_{\rm rot}^2$.}
\end{figure}

\begin{figure}
\includegraphics{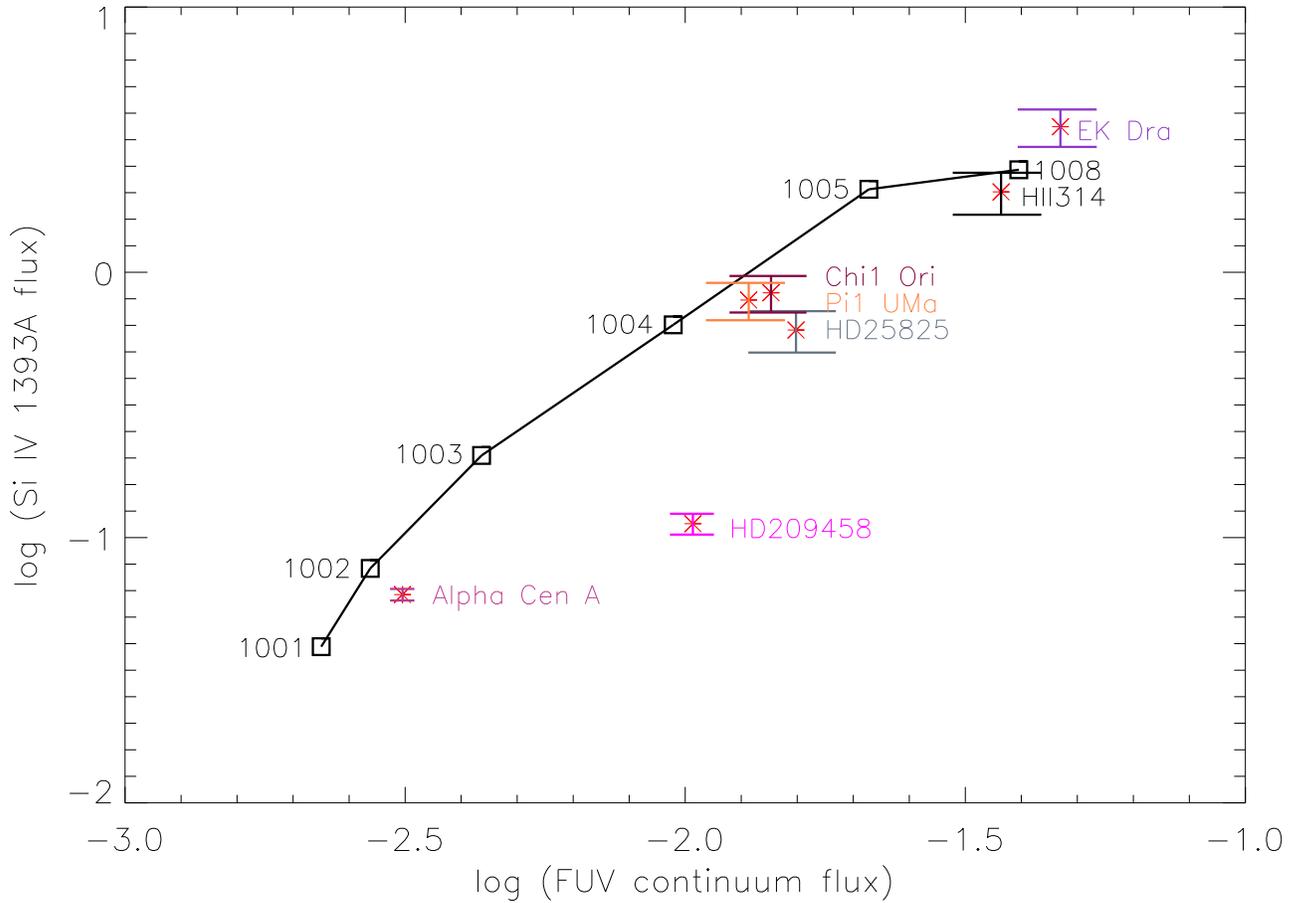}
\caption{Asterix symbols are for the observed Si~IV 1393~\AA\ stellar fluxes 
plotted vs stellar 1382--1392~\AA\ continuum fluxes both at a distance of 
1~AU. The individual stars are identified. The square symbols plot the
predicted Si~IV 1393~\AA\ flux vs 1398.6~\AA\ continuum flux of the 
\citet{Fontenla2011}
models also at a distance of 1~AU. The models numbers are marked.}
\end{figure}

\begin{deluxetable}{lccccccccc}
\tablewidth{0pt}
\tabletypesize{\scriptsize}
\rotate
\tablenum{1}
\tablecaption{Stellar Properties and Observing Log}
\label{tab:log}
\tablehead{\colhead{Star} & \colhead{$P_{\rm rot}$(days)\tablenotemark{a}} & 
\colhead{Age($10^9$ yr)} & \colhead{d(pc)\tablenotemark{b}} &
\colhead{$R/R_{\odot}$\tablenotemark{c}} &
\colhead{Instrument} & \colhead{Dataset} & 
\colhead{Grating} & \colhead{Central $\lambda$} &  \colhead{T$_{\rm exp}$(s)}}
\startdata
HII314 & 1.47 & 0.1 & $135.8\pm 3.0$ & $1.0\pm0.10$ & COS & LB6401010  & 
G130M & 1309 &  ~814.1\\
HII314 & 1.47 & 0.1 & $135.8\pm 3.0$ & $1.0\pm0.10$ & COS & LB6401020  & 
G130M & 1318 &  ~990.1\\
HII314 & 1.47 & 0.1 & $135.8\pm 3.0$ & $1.0\pm0.10$ & COS & LB6401030  & 
G130M & 1291 &  1402.0\\
HII314 & 1.47 & 0.1 & $135.8\pm 3.0$ & $1.0\pm0.10$ & COS & LB6401040  & 
G130M & 1300 &  1405.2\\
EK Dra & 2.60 & 0.03--0.05 & $34.1\pm 0.4$ & $0.95\pm 0.10$ & COS & 
LB3E34010  & G130M & 1291 & 1160.0\\
$\pi^1$ UMa & 4.89 & 0.3 & $14.36\pm 0.08$ & $0.95\pm 0.10$ & COS & 
LB3E26010 & G130M & 1291 & 1300.4\\
$\chi^1$ Ori & 5.10 & 0.3 & $8.66\pm 0.02$ & $0.96\pm 0.10$ & COS & 
LB3E06010& G130M & 1291 & 1300.4\\
HD25825 & 7.3 & 0.6 & $45.9\pm 2.7$ & $1.0\pm 0.10$ & COS & 
LB3E41010  & G130M & 1291 &  1160.0\\
HD209458 & 11.4 & $4\pm 2$ & $49.6\pm 2.0$ & $1.125\pm 0.023$ & COS & 
LB4M\tablenotemark{b} & G130M & & 22441.0\\
$\alpha$ Cen A & 28 & 3.9--4.9 & $1.325\pm 0.007$ & $1.224\pm 0.003$ & STIS & 
   & E140H & 1234 & 4695.2\\
$\alpha$ Cen A & 28 & 3.9--4.9 & $1.325\pm 0.007$ & $1.224\pm 0.003$ & STIS & 
   & E140H & 1416 & 4695.2\\
Solar flux & 25.4 & 4.6 &   &    &  SORCE  &  &       &      &  \\
 
\enddata
\tablenotetext{a}{Rotation Periods: HII314 \citep{Rice2001}, 
EK~Dra \citep{Strassmeier1998,Jarvinen2007}, $\pi^1$~UMa \citep{Gaidos2000},
$\chi^1$~Ori \citep{King2005}, 
HD25825 (mean of 4 Hyades stars with B--V = 0.59--0.61 \citep{Radick1987}),
HD209458 \citep{Silva-Valio2008}, $\alpha$~Cen~A \citep{Barnes2007}.}
\tablenotetext{b}{Data from SIMBAD except for HII314 for which we have used 
the \citet{An2007} distance to the Pleiades.}
\tablenotetext{c}{Data from \citet{Ribas2005} except for HD 209458 
\citep{Knutson2007}, $\alpha$~Cen~A \citep{deMeulenaer2010}, and HII314 
\citep{Ayres1999}. Errors for the other stars are generous estimates.}  

\end{deluxetable}

\begin{deluxetable}{lccccccccc}
\tablewidth{0pt}
\tabletypesize{\scriptsize}
\tablenum{2}
\tablecaption{Far-UV Continuum Fluxes and Brightness Temperatures}
\label{tab:profile}
\tablehead{\colhead{Star} & \colhead{Wavelength} & \colhead{Flux} & 
\colhead{Surface} & \colhead{Continuum} & \colhead{Dark} & \colhead{Ratio} & 
\colhead{Measurement} & \colhead{Systematic} & \colhead{T$_{\rm B}$}\\
\colhead{} & \colhead{Range} & \colhead{(cgs)} & 
\colhead{Flux\tablenotemark{a}} & \colhead{Counts} & 
\colhead{Counts} & \colhead{(C/D)} & \colhead{Error(\%)} & 
\colhead{Error(\%)} & \colhead{(K)}} 

\startdata
HII314 & 1158--1171 & 4.13E-17 & $1497\pm 282$ & 257 & 508 & 0.51 & 10.8 & 
15.4 & $5918^{+49}_{-59}$\\
       & 1244--1249 & 5.76E-17 & $2085\pm 374$ & 223 & 195 & 1.14 & 9.2 & 
15.4 &  $5712^{+47}_{-55}$\\
       & 1267--1273 & 3.75E-17 & $1358\pm 259$ & 182 & 234 & 0.78 & 11.2 & 
15.4 & $5515^{+47}_{-56}$\\
       & 1315--1328 & 4.20E-17 & $1520\pm 265$ & 358 & 508 & 0.70 & 8.2 & 
15.4 &  $5382^{+43}_{-50}$\\
       & 1337--1351 & 4.73E-17 & $1713\pm 293$ & 418 & 547 & 0.76 & 7.4 & 
15.4 & $5346^{+42}_{-50}$\\ 
       & 1372--1380 & 4.74E-17 & $1717\pm 317$ & 229 & 312 & 0.73 & 10.2 &
15.4 & $5253^{+45}_{-53}$\\
       & 1382--1392 & 4.67E-17 & $1692\pm 305$ & 273 & 391 & 0.70 & 9.4 & 
15.4 & $5218^{+44}_{-52}$\\
       & 1414--1424 & 6.69E-17 & $2424\pm 420$ & 343 & 391 & 0.88 & 7.9 & 
15.4 & $5224^{+43}_{-51}$\\ 
EK Dra & 1337--1351 & 8.18E-16 & $2071\pm 333$ & 1826& 138 & 13.2 & 2.4 & 
15.9 & $5397^{+41}_{-47}$\\
       & 1382--1392 & 9.45E-16 & $2393\pm 386$ & 1396&  99 & 14.1 & 2.8 & 
15.9 & $5310^{+41}_{-47}$\\
       & 1414--1424 & 1.09E-15 & $2760\pm 446$ & 1406&  99 & 14.2 & 2.8 & 
15.9 & $5260^{+41}_{-48}$\\
$\pi^1$ UMa & 1337--1351 & 1.10E-15& $494\pm 79.2$ & 2752& 155 &17.8 & 2.0 & 
15.9 & $5033^{+35}_{-41}$\\ 
       & 1382--1392 & 1.48E-15 & $665\pm 107$ & 2452& 111 & 22.1 & 2.1 & 
15.9 & $4984^{+36}_{-41}$\\
       & 1414--1424 & 1.93E-15 & $867\pm 139$ & 2792& 111 & 25.2 & 1.9 & 
15.9 & $4962^{+36}_{-42}$\\
$\chi^1$ Ori & 1337--1351 & 3.22E-15 & $515\pm 81.1$ & 8054& 155 & 52.0& 1.1 & 
15.7 & $5043^{+35}_{-40}$\\
       & 1382--1392 & 4.46E-15 & $713\pm 112$ & 7390& 111 & 66.6 & 1.2 & 
15.7 & $5000^{+35}_{-41}$\\
       & 1414--1424 & 6.12E-15 & $979\pm 154$ & 8854& 111 & 79.8 & 1.1 & 
15.7 & $4991^{+36}_{-42}$\\
HD25825& 1337--1351 & 1.27E-16 & $526\pm 93.9$ & 283 & 138 & 2.04 & 7.3 & 
16.3 & $5048^{+39}_{-46}$\\
       & 1382--1392 & 1.76E-16 & $729\pm 130$ & 261 &  99 & 2.63 & 7.3 & 
16.3 & $5006^{+40}_{-47}$\\
       & 1414--1424 & 2.56E-16 & $1060\pm 185$ & 331 &  99 & 3.34 & 6.3 & 
16.3 & $5011^{+40}_{-47}$\\
HD209458
& 1158--1171& 4.00E-17 & $153\pm 15.3$ & 1216  &2469 & 0.49 & 5.0 & 
8.66 & $5335^{+22}_{-24}$\\
        & 1244--1249& 3.24E-17 & $124\pm 13.4$ & 611 & 949 & 0.64 & 6.5 & 
8.66 & $5012^{+22}_{-25}$\\
        & 1268--1274& 2.56E-17 & $97.6\pm 10.8$ & 606 &1139 & 0.53 & 6.9 &
8.66 & $4886^{+22}_{-25}$\\
        & 1277--1287& 3.52E-17 & $134\pm 12.9$ &1374 &1899 & 0.72 & 4.2 &
8.66 & $4920^{+20}_{-22}$\\
        & 1337--1351& 6.28E-17 & $240\pm 21.8$ & 2704 &2658 & 1.02 & 2.7 & 
8.66 & $4868^{+19}_{-21}$\\
        & 1382--1392& 9.86E-17 & $376\pm 33.8$ & 2814 &1899 & 1.48 & 2.4 & 
8.66 & $4851^{+20}_{-21}$\\
        & 1414--1424& 1.32E-16 & $504\pm 45.0$ & 3289 &1899 & 1.73 & 2.2 & 
8.66 & $4833^{+20}_{-21}$\\
        & 1456--1461& 1.58E-16 & $602\pm 53.9$ & 2520 & 949 & 2.66 & 2.3 &
8.66 & $4774^{+20}_{-22}$\\
        & 1487--1494& 1.96E-16 & $750\pm 66.5$ & 3674 &1329 & 2.76 & 1.9 &
8.66 & $4747^{+20}_{-22}$\\
$\alpha$ Cen A\tablenotemark{b} &
          1158--1171  & 3.05E-14 & $70.2\pm 11.5$ & 2023 &12548& 0.16 & 6.0 & 
15.2& $5161^{+33}_{-38}$\\
        & 1244--1249  & 2.47E-14 & $56.9\pm 8.75$ &  4085 &4511 & 0.91 & 2.3 & 
15.2& $4848^{+29}_{-33}$\\
        & 1277--1287  & 2.24E-14 & $51.6\pm 7.88$ &  8766 &8766 & 1.00 & 1.5 & 
15.2& $4722^{+28}_{-33}$\\
        & 1337--1351  & 2.62E-14 & $60.3\pm 2.70$ &  14986&11708& 1.28 & 1.1 & 
4.34& $4580\pm 09$\\
        & 1372--1379.3& 4.17E-14 & $96.0\pm 4.30$ &  12044&5923 & 2.03 & 1.1 & 
4.34& $4589\pm 09$\\
        & 1382--1392  & 4.19E-14 & $96.5\pm 4.30$ &  14506&8104 & 1.79 & 1.0 & 
4.34& $4561\pm 09$\\
        &1396.1--1398.6&4.68E-14 & $109\pm 5.21$  &  3921 &2011 & 1.95 & 2.0 & 
4.34& $4559\pm 09$\\
        & 1414--1424  & 6.03E-14 & $139\pm 6.13$ &   20294&7921 & 2.56 & 0.8 & 
4.34& $4554\pm 09$\\
        & 1426--1431  & 6.10E-14 & $140\pm 6.30$ &   9991 &3934 & 2.54 & 1.2 & 
4.34& $4532\pm 09$\\
        & 1438--1444  & 7.40E-14 & $170\pm 7.57$ &   13542&4680 & 2.89 & 1.0 & 
4.34& $4541\pm 09$\\
        & 1446--1454  & 7.88E-14 & $181\pm 8.02$ &   16266&6201 & 2.62 & 0.9 & 
4.34& $4532\pm 09$\\
        & 1475--1480  & 9.75E-14 & $225\pm 10.0$ &   11987&3422 & 3.50 & 1.0 & 
4.34& $4512\pm 09$\\
        & 1494.5--1500& 1.19E-13 & $274\pm 12.1$ &  14746&4129 & 3.57 & 0.9 & 
4.34& $4507\pm 09$\\
Sun\tablenotemark{c}  & 
          1158--1171  & 2.53E-03 & $117\pm 5.4$ & & & & & 5.0 & $5273\pm 10$\\
        & 1244--1253  & 2.06E-03 & $95.2\pm 4.8$ & & & & & 5.0 & $4949\pm 11$\\
        & 1277--1287  & 1.51E-03 & $69.8\pm 3.6$& & & & & 5.0 & $4783\pm 10$\\
        & 1315--1328  & 1.97E-03 & $90.9\pm 4.6$& & & & & 5.0 & $4724\pm 10$\\
        & 1337--1351  & 1.79E-03 & $82.9\pm 4.3$& & & & & 5.0 & $4643\pm 10$\\
        & 1371--1381  & 2.70E-03 & $125\pm 6.4$ & & & & & 5.0 & $4641\pm 10$\\
        & 1382--1392  & 2.67E-03 & $123\pm 6.3$ & & & & & 5.0 & $4610\pm 10$\\
        & 1396--1399  & 3.10E-03 & $143\pm 7.4$ & & & & & 5.0 & $4614\pm 10$\\
        & 1408--1422  & 3.70E-03 & $171\pm 8.7$ & & & & & 5.0 & $4607\pm 10$\\
        & 1414--1424  & 3.64E-03 & $168\pm 8.5$ & & & & & 5.0 & $4593\pm 10$\\
        & 1426--1431  & 3.72E-03 & $172\pm 8.8$ & & & & & 5.0 & $4574\pm 11$\\
        & 1438--1444  & 4.37E-03 & $202\pm10.3$ & & & & & 5.0 & $4577\pm 11$\\
        & 1446--1454  & 4.51E-03 & $208\pm10.6$ & & & & & 5.0 & $4561\pm 11$\\
        & 1475--1480  & 5.83E-03 & $270\pm13.4$ & & & & & 5.0 & $4550\pm 11$\\
        & 1494.5--1500& 6.89E-03 & $318\pm16.3$ & & & & & 5.0 & $4539\pm 11$\\

\enddata 
\tablenotetext{a}{Surface fluxes (units: ergs cm$^{-2}$ s$^{-1}$ \AA$^{-1}$) 
computed using the stellar distances and radii listed in Table~1. Flux errors 
are computed from a quadradic sum of the measurement and systematic errors.}
\tablenotetext{b}{Observed with the STIS E140H grating.}
\tablenotetext{c}{Observed by SORCE.}
\end{deluxetable}

\end{document}